\documentclass[largeformat]{interact}

\usepackage{epstopdf}
\usepackage[caption=false]{subfig}
\usepackage[colorlinks=true,linkcolor=black,anchorcolor=black,citecolor=black,filecolor=black,menucolor=black,runcolor=black,urlcolor=black]{hyperref}

\newif\ifshowauthors
\showauthorstrue
\ifshowauthors
  \usepackage[authoryear]{natbib}
  \setcitestyle{aysep={}}
  \renewcommand\bibfont{\fontsize{10}{12}\selectfont}
\else
  \usepackage[numbers,sort&compress]{natbib}
  \bibpunct[, ]{[}{]}{,}{n}{,}{,}
  \renewcommand\bibfont{\fontsize{10}{12}\selectfont}
  \makeatletter
  \def\NAT@def@citea{\def\@citea{\NAT@separator}}
  \makeatother
\fi

\theoremstyle{plain}

\theoremstyle{definition}

\theoremstyle{remark}

\usepackage{wasysym}  
\usepackage{bm}       
\usepackage{wrapfig}  

\newcommand\mnras{MNRAS}
\newcommand\apj{ApJ}
\newcommand\prl{Phys. Rev. Lett}
\newcommand\aap{A\&A}

\newcommand\Msun{\text{M}_{\astrosun}} 

\begin{document}


\title{\vspace{-.2cm}Supermassive Black Holes in the Early Universe}

\author{
\name{Aaron Smith\textsuperscript{a}\thanks{Corresponding author. Email: \href{mailto:asmith@astro.as.utexas.edu}{arsmith@mit.edu}} and Volker Bromm\textsuperscript{b}}
\affil{\textsuperscript{a}Department of Physics,
  Massachusetts Institute of Technology, Cambridge, MA 02139, USA; \\ \textsuperscript{b}Department of Astronomy, The University of Texas at Austin, Austin, TX 78712, USA}
}

\maketitle

\begin{abstract}
The emergence of the first black holes during the first billion years of cosmic history marks a key event in cosmology. Their formation is part of the overall process of ending the cosmic dark ages, when the first stars appeared in low-mass dark matter haloes about a few 100 million years after the Big Bang. The first stars, galaxies, and black holes transformed the Universe from its simple initial state into one of ever increasing complexity. We review recent progress on our emerging theoretical picture of how the first black holes appeared on the cosmic scene, and how they impacted the subsequent history of the Universe. Our focus is on supermassive black holes, in particular assessing possible pathways to the formation of the billion-solar-mass black holes inferred to power the luminous quasars at high redshifts. We conclude with a discussion of upcoming empirical probes, such as the \textit{James Webb Space Telescope} (\textit{JWST}), and the Laser Interferometer Space Antenna (LISA), further ahead in time.
\end{abstract}

\begin{keywords}
black holes; galaxies: formation; galaxies: evolution; galaxies: high-redshift
\end{keywords}

\section{Introduction}
\label{sec:intro}
One of the key questions in modern cosmology is to understand when black holes first appeared on the cosmic scene, and when they began to impact the evolution of the Universe. Black holes are extreme objects, left behind when massive stars die, with a gravitational force that is so strong that nothing can escape, not even light \citep{Chandrasekhar1974}. The first black holes in particular played an important part in ending the cosmic dark ages, the period in the history of the Universe before any stars or galaxies had formed \citep{BrommYoshida2011,LoebFurlanetto2013}. The dark matter, the dominant form of mass in the Universe, was initially distributed very uniformly, but was endowed with small fluctuations in its density, which in turn originated in quantum processes in the moments briefly after the Big Bang. These primordial perturbations were slowly amplified by gravity, to the point where some regions of dark matter could break away from the overall expansion of the Universe. Held together by their own gravity, those regions gave rise to roughly spherical dark matter ``haloes'', which also contained trace amounts of normal matter, the hydrogen and helium gas made in the Big Bang.

Some haloes were able to host the first stars and galaxies, formed out of primordial gas. Within the standard model of how structures formed in the Universe, assuming the presence of dark matter and dark energy, this epoch of first light is predicted to have taken place a few $100$~Myr after the Big Bang. In cosmology, it is convenient to express ages in terms of the ``redshift'', denoted by the symbol $z$, which measures how much the wavelength of light has been stretched by the expanding Universe for present-day observers. Thus, a larger redshift implies earlier times, closer to the Big Bang, such that the time when the first stars formed corresponds to $z \sim 20$--$30$. This period marks the fundamental transition from the simple initial conditions in the very early Universe to the ever increasing complexity that we see around us today.

\begin{table}
\tbl{Summary of acronyms used throughout the review.}
{\begin{tabular}{ll} \toprule
  Acronym & Explanation \\ \midrule
  21\,cm & Hyperfine-structure transition of neutral hydrogen \\
  AGN & Active galactic nuclei (galaxy centres with accreting SMBHs) \\
  CMB & Cosmic microwave background (created $\sim380,000$\,yr after the Big Bang) \\
  DCBH & Direct-collapse black hole (massive seed without fragmentation) \\
  $\text{H}_2$ & Molecular hydrogen (primary coolant at low temperatures) \\
  HMXB & High-mass X-ray binary (black hole accreting from companion star) \\
  IGM & Intergalactic medium (low density gas between galaxies) \\
  \textit{JWST} & \textit{James Webb Space Telescope} (infrared wavelengths) \\
  LIGO & Laser Interferometer Gravitational-wave Observatory \\
  LISA & Laser Interferometer Space Antenna (gravitational-waves) \\
  LW & Non-ionizing (soft-UV) Lyman-Werner photons \\
  Ly$\alpha$ & Lyman-alpha line of atomic hydrogen ($2p\rightarrow1s$ transition) \\
  Pop~III & Primordial first stars (formed a few 100~Myr after the Big Bang) \\
  SKA & Square Kilometre Array (radio interferometer) \\
  SMBH & Supermassive black hole ($M_\bullet \gtrsim 10^6\,\Msun$) \\
 \bottomrule
\end{tabular}}
\label{tab:acronyms}
\end{table}

From a theoretical viewpoint, the formation of the first stars marks the moment when differences in density became increasingly pronounced. Previously, the growth of the small density perturbations was governed by simple linear equations, describing a Gaussian random field, with parameters calibrated to exquisite precision by measurements of the cosmic microwave background (CMB). Here, key probes of the CMB were the satellite-borne \textit{Wilkinson Microwave Anisotropy Probe} (\textit{WMAP}) and \textit{Planck} missions. Afterwards, studying the emerging structures requires sophisticated numerical simulations, tracing both the dark matter and baryonic (gaseous) components. The first stars\footnote{The first stars, formed out of primordial hydrogen and helium gas, are often called Population~III (Pop~III) stars. This terminology extends the traditional sequence of astronomy, where metal-rich stars, like our Sun, are Population~I stars, and the metal-poor stars in the outer regions of the Milky Way constitute Population~II.} also initiated the transition from primordial gas to one enriched with heavy chemical elements (``metals'' in the parlance of astronomy), thus providing more efficient cooling channels that were not accessible in the early Universe \citep{Glover2013,Barger2017}.

The focus of this review is on supermassive black holes (SMBHs) in the early Universe, with masses\footnote{In astronomy, the mass of the Sun is denoted with $\Msun \approx 2 \times 10^{30}$\,kg.} of $M_{\bullet} \gtrsim 10^6\,\Msun$, but it is likely that the first black holes were less massive, emerging from the collapse of Pop~III stars \citep{MadauRees2001}. Such stellar remnants would have masses of a few $\sim 10\,\Msun$. A fraction of these stellar-mass black holes might give rise to a population of tightly-bound double stars, called high-mass X-ray binaries (HMXBs), where the black hole is in orbit around a normal star. These sources release copious amounts of high-energy (X-ray) radiation when gas from the companion star is falling onto the black hole in the binary system, experiencing ferocious frictional heating in the process. This heat energy could in turn significantly influence the early intergalactic medium (IGM), through widespread X-ray ionization and heating \citep{Jeon2012,Jeon2014}.

Another intriguing class of black holes may have formed in the ultra-early Universe, in the chaotic conditions that prevailed briefly after the Big Bang. Typically, such ``primordial black holes'' are predicted to have masses that are much smaller than stars, extending all the way down to the Planck mass, $m_\text{P} \sim 10^{-8}\,\text{kg}$, the quantum-gravitational unification scale\footnote{Intuitively, $m_\text{P}$ is the mass of a black hole that is just massive enough to gravitationally hold on to its mass, before quantum-tunneling is able to return matter to the outside of the black hole event horizon. Classically, nothing is allowed to do that.}. Recently, a variant of primordial black holes has been proposed as the origin for the elusive dark matter \citep[][]{Bird2016}. To account for the dark matter, a sub-class of primordial black holes could have formed with typical masses of a few tens of solar masses, in line with the masses of the binary black hole mergers observed by the Laser Interferometer Gravitational-wave Observatory (LIGO). Since these objects would have formed before the epoch of Big Bang nucleosynthesis, when the primordial hydrogen and helium was produced, they would not have shown up in the baryon budget encapsulated in the abundance ratios of the primordial chemical species\footnote{Standard analysis of the primordial chemistry indicates that the dark matter has to be ``non-baryonic'', made up of a new kind of elementary particle.}.

We know that the Universe was able to form SMBHs with $M_{\bullet} \gtrsim 10^9\,\Msun$ in the centers of giant galaxies. Spectacular confirmation has recently come from the direct imaging of the immediate vicinity of such an object in the Messier~87 (M87) galaxy with the Event Horizon Telescope. This telescope consists of radio antennae that are distributed all over the Earth, thus providing the incredible power to resolve the central black hole in M87. Furthermore, observations of luminous, extremely distant quasars imply that SMBHs were already in place at very early times, corresponding to redshifts $z\gtrsim 7$ \citep[][]{Fan2006,Mortlock2011,Wu2015,Banados2018}. Quasars are vastly scaled-up cousins to the HMXB sources mentioned above, where again huge amounts of heat energy are released when surrounding gas falls onto the central black hole.

The question then arises how lower-mass black holes, acting as seeds, could have grown into these massive structures in the limited time available. Traditionally, stellar-mass seeds were assumed, possibly the remnants of massive Pop~III stars. Those seeds with up to $M_{\bullet} \sim 100\,\Msun$ could then grow through mass assembly, or ``accretion'', to reach the SMBH target mass by $z\sim 7$, as required by the observations \citep{HaimanLoeb2001}. There is a limit to the rate of how fast this assembly could have occurred, first worked out by the British astrophysicist Arthur Eddington. The idea here is that the higher the mass assembly rate, the larger the output of radiation. At some point the outward pressure exerted by the radiation is able to overwhelm the inward force of gravity. This balancing of forces defines the Eddington (accretion) rate, the maximum growth rate of black holes.

The stellar-seed scenario has been challenged, however, when taking into account the effects of energy input from stars on the efficiency of the accretion flows \citep[][]{JohnsonBromm2007,Milosavljevic2009}. Specifically, the ultra-violet (UV) radiation from the massive black hole progenitor stars will heat the surrounding gas, thus driving strong outflows. As a consequence, the newly formed black hole finds itself, at least initially, in a low-density environment, taking time to replenish the gas supply in its vicinity. Mass growth rates, therefore, are greatly suppressed below the Eddington rate for about the age of the Universe at that time ($\sim100$\,Myr). Under such conditions, stellar seeds require growth rates that are somehow able to exceed the Eddington rate to reach the SMBH target mass \citep[][]{Jeon2014}.

In response to this ``timing crisis'', the alternative scenario of direct-collapse black holes (DCBHs) has been suggested \citep{Rees1984,BrommLoeb2003}. The idea here is to trigger the collapse of a primordial gas cloud into a so-called ``atomic cooling halo'', whose gravitational potential well is deep enough to heat the infalling gas to temperatures of $T \gtrsim 10^4\,\text{K}$. Collisions between gas particles are thus sufficiently energetic to excite hydrogen atoms from the ground state. The subsequent recombination radiation is then able to radiate away the heat energy, generated during the compression of the gas. The strongest cooling channel is often provided by the atomic hydrogen Lyman-alpha (Ly$\alpha$) line, when the atom returns from its first excited state into the ground state. To activate this ``atomic cooling'' mechanism, the dark matter halo has to have a mass of  $M_\text{halo} \sim 10^8\,\Msun$ at redshifts $z \gtrsim 10$.

Normally, such collapse would be accompanied by vigorous fragmentation, and thus star formation. The star-forming activity in turn would trigger strong energy input, which would act to dispel the gas supply from the host halo. To accomplish the assembly of a significant fraction of the gas into a supermassive object at the halo center, the DCBH model assumes that any fragmentation is suppressed by preventing the gas from cooling down. Molecular hydrogen ($\text{H}_2$), different from the atomic hydrogen in Ly$\alpha$ emission, would enable such low-temperature cooling, because of the low-energy rotational and vibrational quantum states that can be excited. Destroying the H$_2$ may be possible if the primordial cloud is suffused in a very strong flux of soft-UV photons with energies just below the threshold to ionize hydrogen, but capable of dissociating molecular hydrogen\footnote{These photons reside in the so-called Lyman-Werner (LW) wavelength band, giving rise to the term LW radiation.}. Recent cosmological simulations have confirmed that such ``collapse without fragmentation'' can indeed lead to the build-up of massive central structures, as we will discuss further below.

A population of black holes in the early Universe would result in the build-up of a cosmic X-ray background, a near-uniform, pervasive radiation field in the X-ray band. Of particular importance might be the HMXB sources mentioned above, possibly resulting from the first stars \citep[][]{Mirabel2011}. Their emission could contribute to the widespread ionization and heating of the primordial gas, given the long mean free paths of the high-energy X-ray photons \citep[][]{KuhlenMadau2005}. Prior to what cosmologists call the ``epoch of reionization'' about a billion years after the Big Bang, the intergalactic gas was still substantially neutral, but any degree of (partial) ionization from the X-ray background would impact the primordial chemistry, in particular the pathways towards forming $\text{H}_2$, the main coolant in low-temperature metal-free gas \citep[][]{GalliPalla2013}.

Another important thermal effect of an early X-ray background may be the way it affects the coupling of the hyperfine-structure levels of neutral hydrogen to the kinetic energy of the cosmic gas. Those levels in turn determine the strength of the 21\,cm spin-flip transition, which could be observed with dedicated radio telescopes \citep[][]{Furlanetto2006,PritchardLoeb2012}. Understanding the detailed physics behind this coupling is a key challenge in the emerging field of 21\,cm cosmology, with pathfinder experiments such as the Experiment to Detect the Global Epoch of Reionization Signature (EDGES) already underway \citep{Bowman2018}, and with the large-scale Square Kilometre Array (SKA) on the horizon.

We have a number of tantalizing windows from observations into the properties and impact of black holes at high redshifts. Next to the constraints from the most distant, hyper-luminous quasars on the growth of the first SMBHs, there are hints of a significant correlation between the unresolved cosmic X-ray and (near-) infrared backgrounds, possibly indicating a physically related origin, such as a population of massive Pop~III stars \citep{Cappelluti2013}. Another promising probe of the first stellar-mass black holes are the gravitational waves emitted by binary black hole mergers. Estimates indicate that about 1\% of the black hole merger events detected by LIGO could originate in Pop~III progenitors \citep[][]{Hartwig2016}. In the near future, the direct hunt for the first SMBHs will be carried out by the \textit{James Webb Space Telescope} (\textit{JWST}), with a particular focus on searching for DCBH sources. Once we have identified such promising SMBH candidates with the \textit{JWST}, deep spectroscopic follow-up with the upcoming suite of extremely large telescopes on the ground will provide key diagnostics for the physical nature of these \textit{titans of the early Universe} \citep[see][]{Woods2019}.

The outline for this review is as follows. We begin with a discussion of the underlying black hole formation physics, within the context of standard cosmology, covering the entire range of possible masses. We next address the multiple ways in which black holes can impact the early IGM, acting both as drivers of cosmic evolution and providing empirical probes of the sources at work here. After assessing the potential for such probes in testing our emerging theoretical picture, we will conclude with our outlook for the future. Many of these themes consistently appear in reviews of this subject, albeit with particular emphases and continuous refinements. For further exploration, we refer the reader to  this rapidly expanding literature \citep[][]{Volonteri2012,Haiman2013,JohnsonHaardt2016,LatifFerrara2016,Woods2019}.

\begin{figure}
\centering
\includegraphics[width=0.7\textwidth]{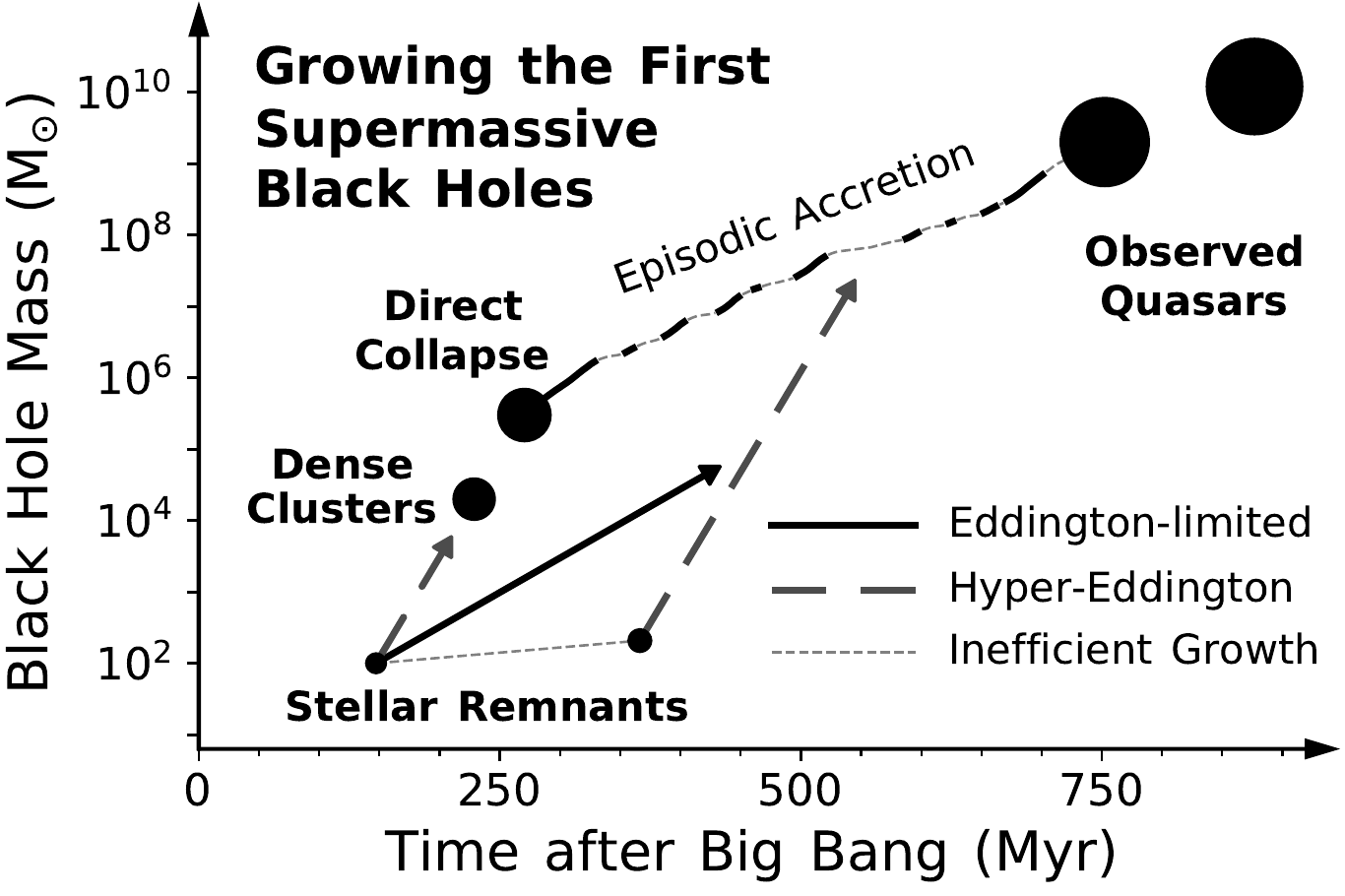}
\caption{Illustrating the timing crisis for supermassive black hole growth in the early Universe. The observed quasars with masses of $M_\bullet \approx 10^9$--$10^{10}\,\Msun$ at redshifts of $z \approx 6$--$7$ most likely originated from the following scenarios: the direct-collapse of a primordial gas cloud for which cooling does not initiate fragmentation, dense star clusters in which runaway collisions trigger black hole formation, and stellar-remnant black holes experiencing hyper-Eddington mass growth. The subsequent growth is dominated by episodic galaxy mergers and infall from streams of cold gas along dark matter filaments \citep[adapted from][]{SmithAG2017}.} \label{fig1:Smith2017}
\end{figure}

\section{Formation}
\label{sec:formation}
There are several pathways to forming astrophysical black holes, including both canonical and exotic scenarios. We summarize the most relevant ones for the high-redshift Universe with the diagram in Figure~\ref{fig1:Smith2017} with further details in the following subsections. This roadmap provides the context for understanding the challenges encountered in order to grow efficiently and make connections with observational studies, especially in the context of the aforementioned quasar-seed timing problem. Early simulations of collapsing primordial gas clouds showed that most of the gas fragments into dense stellar clumps which eventually virialize into a spheroidal galactic bulge \citep{LoebRasio1994}. Even without fragmentation, conservation of angular momentum during collapse generates a centrifugal barrier inhibiting rapid mass accretion \citep{EisensteinLoeb1995,Sugimura2018}. However, the presence of a central SMBH of mass $M_\bullet \gtrsim 10^6\,\Msun$ would stabilize the inner region of the disc and allow growth by steady accretion as long as the gas supply remains \citep{Li2007}. Thus, the crucial bottleneck is the formation of the initial SMBH, such that rapid seeding or growth is unavoidable within our current understanding of the most distant quasars.

\subsection{Primordial black holes}
\label{sec:primordial}
The very first black holes may be byproducts of the extreme conditions of the ultra-early Universe. Specifically, the high pressure and energy fields seeded by inhomogeneous quantum fluctuations could have triggered sufficient compression of overdense regions to give rise to primordial black holes \citep{CarrHawking1974}. These could be Planck relics with initial masses from $m_\text{P} \sim 10^{-8}\,\text{kg}$ upwards, experiencing highly inefficient growth throughout cosmic history with the possible exception of a rapid growth mode in dense galactic centres. Interestingly, miniature primordial black holes with masses $M_\bullet \lesssim 10^{-19}\,\Msun \approx 10^{11}\,\text{kg}$ are of little astrophysical consequence as the evaporation timescale due to Hawking radiation\footnote{Hawking radiation, named after the British theoretical physicist Stephen Hawking, is black-body radiation predicted to be released by black holes due to quantum effects near the event horizon.} becomes significantly less than the age of the Universe. More massive primordial black holes would persist to the present day and have been invoked as plausible dark matter candidates. However, to account for the inferred dark matter budget, e.g. from measurements of the CMB \citep{Planck2016}, the primordial black hole mass is restricted to $20\,\Msun \lesssim M_\bullet \lesssim 100\,\Msun$, coincident with the masses of recently discovered coalescing binary black holes \citep{LIGO2016,Bird2016}. Lower masses are excluded by microlensing surveys \citep{Alcock2001}, while high masses would disrupt wide binaries \citep{Yoo2004}. Still, primordial black holes could exist outside this range at much lower abundances, thereby providing an alternative seeding mechanism for SMBHs.

\subsection{Stellar remnant black holes}
\label{sec:stellar}
The first black holes to form via conventional gravitational collapse were remnants of the first (Pop~III) stars. Due to the substantially lower metal content, compared to present-day star forming clouds, these stars likely exhibited a top-heavy initial mass function \citep{BrommYoshida2011}. The most massive stars ($\gtrsim 10\,\Msun$) underwent efficient nuclear fusion and had short main sequence lifetimes ($\lesssim 3\,\text{Myr}$) before their eventual deaths as either core-collapse supernovae ($10$--$40\,\Msun$), superluminous pair-instability supernovae ($140$--$260\,\Msun$), or by directly collapsing to a black hole \citep[$40$--$140\,\Msun$, $\gtrsim260\,\Msun$,][]{HegerWoosley2002,Heger2003}. However, the first stellar remnant black holes were ``born starving'', as their strong ionizing feedback was capable of evaporating the surrounding gas \citep{Whalen2004,JohnsonBromm2007}. Furthermore, without a stabilizing galactic potential well they were susceptible to ejection from their host halos by natal kicks imparted during formation, precluding further growth \citep{WhalenFryer2012,SmithRegan2018}. The subsequent self-enrichment of metals from stochastic supernovae and external pollution from more evolved neighbouring galaxies prompted a transition to a Pop~II mode of star formation with enhanced fragmentation, establishing a universal low-mass dominated initial mass function \citep{SafranekShrader2014,SmithWise2015,Jeon2017}. This marked an end to the formation of isolated massive stellar remnant black hole seeds.

\begin{figure}
\centering
\includegraphics[width=0.7\textwidth]{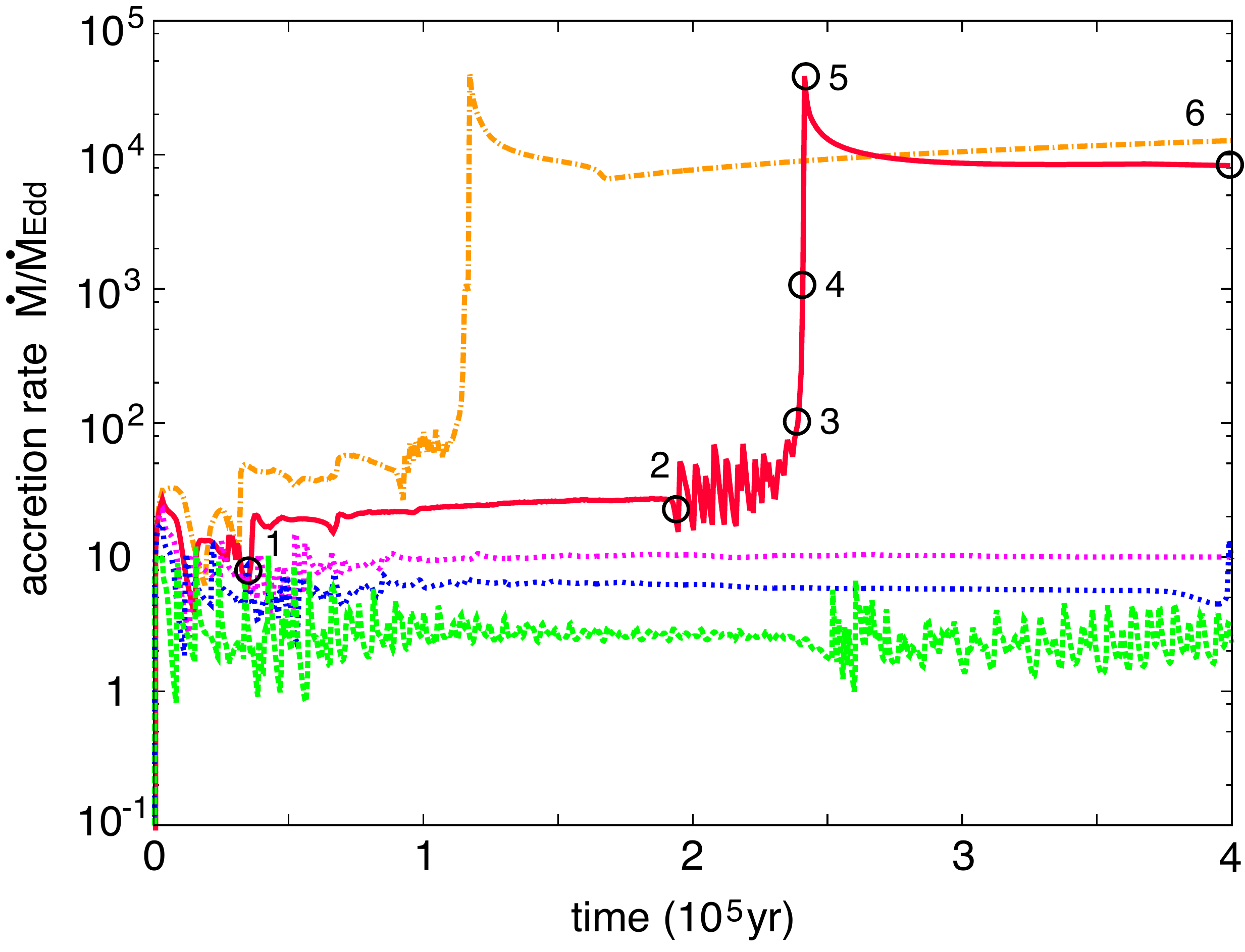}
\caption{Accretion rate history from one-dimensional radiation-hydrodynamics simulations of black holes with mass $M_{\bullet} = \{1,3,5,10,20\} \times 10^3\,\Msun$ (green, blue, magenta, red, and orange, respectively). The radiative efficiency is based on the photon trapping model and the ambient gas density is $n_\infty = 10^5\,\text{cm}^{-3}$. The higher black hole masses ($M_{\bullet} \gtrsim 10^4\,\Msun$) exhibit a significant jump in their accretion rates, which maintains an approximately constant super-Eddington value as the ionized region is confined within the Bondi radius \citep[adopted from][]{Inayoshi2016}.} \label{fig2:Inayoshi2016}
\end{figure}

\subsection{Eddington-limited accretion}
\label{sec:accretion}
Once a black hole has formed it may grow by accreting gas from the surrounding medium. To be captured a particle must fall within the causal event horizon, i.e. the Schwarzschild radius $R_\text{S} = 2 G M_\bullet / c^2$. However, under the assumption of spherical inflow all gas is susceptible to accretion if it falls within the so-called Bondi-Hoyle radius $R_\text{B} = 2 G M_\bullet / c_\text{s}^2$, where $c_\text{s}$ denotes the sound speed \citep{Bondi1952}. This model forms a benchmark for understanding black hole growth. The accretion rate is then regulated by the build-up of pressure from the radiation emitted by the infalling gas as it encounters viscous heating. The Eddington limit is the maximal luminosity an accreting object can achieve before launching an intense radiation-driven wind. Under spherical accretion the classical Eddington luminosity including only electron scattering is $L_\text{Edd} = 4 \pi G M_\bullet m_\text{p} c / \sigma_\text{T} \sim 10^{38}\,(M_\bullet/\Msun)\,\text{erg/s}$,\footnote{Astronomers often use ``erg'' for units of energy, which translates to SI units as $1\,\text{erg/s} = 10^{-7}\,\text{W}$.} where $m_\text{p}$ is the proton mass and $\sigma_\text{T}$ the Thomson scattering cross-section of free electrons. This limit is independent of the distance from the black hole because both the radiative and gravitational forces fall off as $1/r^2$. Furthermore, black hole energy production is highly efficient compared to other processes, e.g. nuclear fusion, with the observed luminosity conventionally given as $L = \epsilon \dot{M}_\bullet c^2$, where $\epsilon \sim 0.1$. This provides a constraint on SMBH growth rates, such that $\dot{M}_\bullet \propto M_\bullet$ maximally allows exponential growth.

The Eddington limit serves as the canonical criterion for inhibiting black hole growth. However, several complicating factors can alter this simple picture. For example, thin-disc configurations allow radiation to preferentially escape in the polar directions, such that the funneled accretion process in three-dimensions remains undisturbed.
The radial advection velocity in a viscous disc is given by $v_r \sim \alpha c_\text{s} H / r$, where $\alpha$ is the viscous parameter ($<1$) and $H$ is the disc scale height ($<r$). Thus, the high entropy gas inflows are generally subsonic $v_r \lesssim c_\text{s}$, but transport energy faster than photon-diffusion due to the high optical depths.\footnote{``Optical depth'' is a term that describes the (logarithmic) transmission of light through a material, and is calculated along ray segments as $\tau_e = \int_0^L n_e \sigma_\text{T}\,\text{d}\ell$, where $n_e$ denotes the number of free electrons per unit volume. \label{footnote:optical_depth}} Furthermore, other forms of radiation can provide additional feedback, porosity can generate radial stratification, and energy can be lost to the black hole itself. Fortuitous circumstances can alleviate some of the growth restrictions, and in fact a super-Eddington quasar at $z = 6.6$ has recently been discovered in support of robust SMBH growth in the high-redshift Universe \citep{Tang2019}.

\begin{figure}
\centering
\includegraphics[width=0.85\textwidth]{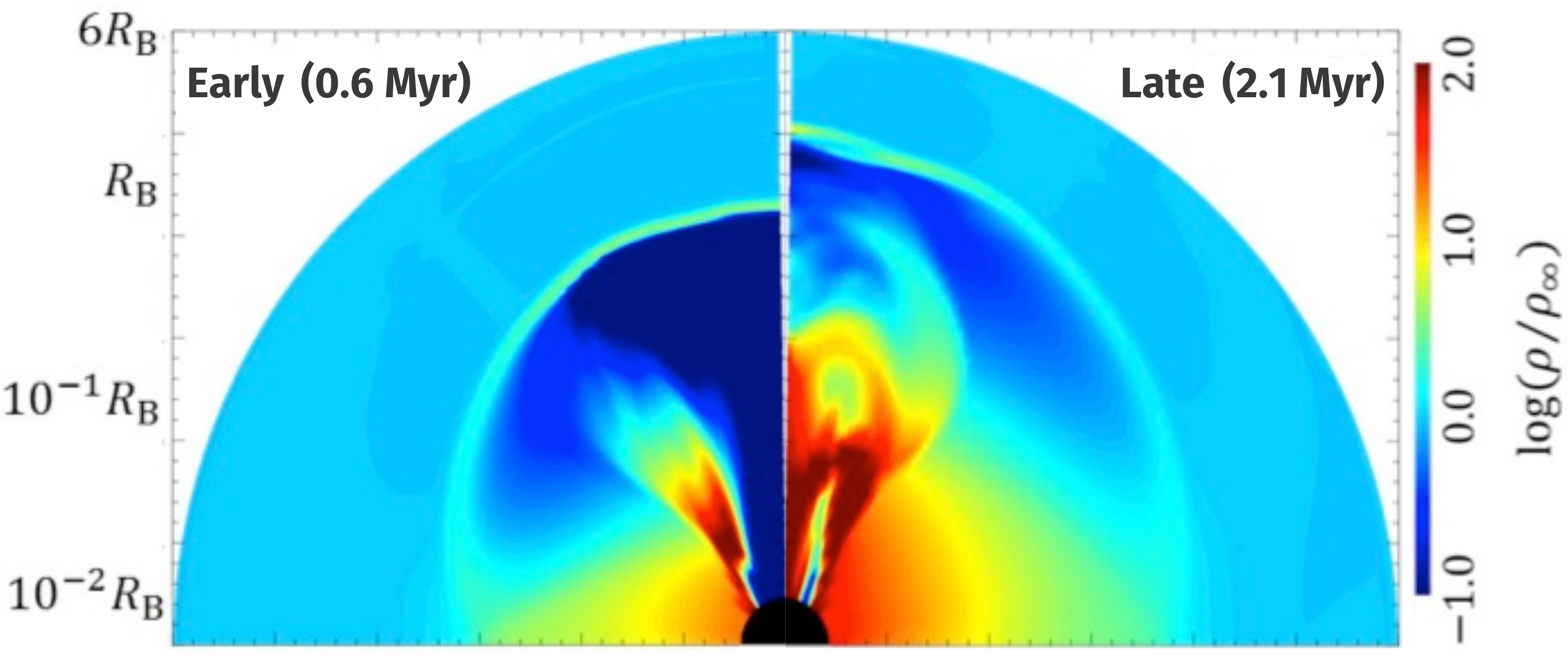}
\caption{Two-dimensional radiation-hydrodynamics simulation demonstrating a transition to super-Eddington accretion rates. The black hole has a mass of $M_{\bullet} \approx 5 \times 10^5\,\Msun$ and ambient density of $n_\infty = 10^5\,\text{cm}^{-3}$. The left and right panels show the gas density during an early stage at $t = 0.6\,\text{Myr}$ ($\approx 0.14\,t_\text{dyn}$) and late stage at $t = 2.1\,\text{Myr}$ ($\approx 0.49\,t_\text{dyn}$), respectively \citep[adapted from][]{Takeo2018}.} \label{fig3:Takeo2018}
\end{figure}

Still, black hole accretion in typical galactic environments is episodic as a result of self-regulating radiative feedback which yields accretion rates that are sub-Eddington when averaged over multiple duty cycles \citep{JohnsonBromm2007,Milosavljevic2009,Park2011}. Maintaining super-Eddington accretion is possible when the black hole is embedded within sufficiently dense gas, which renders the radiation pressure less effective \citep{Begelman1979}. Analytic models using the photon trapping criterion (in which photon diffusion is slower than $v_r$) allow accretion levels far in excess of the Eddington rate \citep{WyitheLoeb2012}. However, such models do not account for photo-heating (hydrodynamical) feedback on the Bondi scale, which could strongly suppress the inflow rate from larger scales \citep{Pacucci2015}. The necessary condition for hyper-Eddington accretion is to overcome photo-heating feedback, which can only happen if the ionized region is confined within the Bondi radius, so multi-frequency radiation hydrodynamics coupling is required. In such one- and two-dimensional radiation hydrodynamics simulations (see Figs.~\ref{fig2:Inayoshi2016}~and~\ref{fig3:Takeo2018}), accretion rates can exceed $\dot{M}_\bullet \gtrsim 10^3\,L_\text{Edd}/c^2$ when the following condition is satisfied: $(M_\bullet/10^5\,\Msun) (n_\infty/10^4\,\text{cm}^{-3}) \gtrsim 2.4\,(\langle \epsilon \rangle / 100\,\text{eV})^{-5/9}$, where $n_\infty$ is the density of the ambient gas and $\langle \epsilon \rangle$ is the mean energy of the ionizing radiation from the central black hole \citep{Inayoshi2016,Takeo2019}. It remains an open question whether super-Eddington growth rates are sustainable considering the violent assembly environments of the first galaxies where newly formed stars and supernovae blow away the surrounding gas.

\subsection{Direct-collapse black holes}
\label{sec:DCBH}
On the other hand, scenarios involving higher initial black hole masses offer promising alternatives to circumvent the quasar-seed timing problem. Primordial galaxies with a delayed onset of star formation facilitate a mechanism to produce massive ($10^4$--$10^6\,\Msun$) black hole seeds, which form ``in one go'' from gas clouds with inefficient cooling. Specifically, according to the Jeans criterion for triggering gravitational instability,\footnote{The Jeans instability causes gas clouds to compress and fragment when the internal pressure is not strong enough to prevent gravitational collapse. This occurs when the free-fall time, $t_\text{ff} \approx 1 / \sqrt{G \rho}$, is shorter than the sound-crossing time, $t_s \approx R / c_s$, with $G$ the universal gravitational constant, $\rho$ the density, $R$ the radius, and $c_s$ the sound speed.} if such a cloud maintains high thermal pressure support then fragmentation is suppressed and the central protostellar core results in a single supermassive star as a result of the significantly altered evolutionary track through density--temperature phase space during collapse (higher thermal pressure add stability). The DCBH scenario depends heavily on the absence of molecular hydrogen, heavy elements, and dust as radiative cooling would lower the temperature and Jeans mass scale. Therefore, DCBHs can only form early in cosmic history under a specific set of rare conditions.

The canonical mechanism to destroy $\text{H}_2$ molecules is by photodissociation from non-ionizing Lyman-Werner photons (``LW'', with energies between 11.2--13.6\,eV) emitted by the first stars \citep{Schaerer2002} and DCBHs \citep{Barrow2018}. LW radiation can travel great distances and build up a LW background modulated by strong local fluctuations that entirely suppresses star formation in minihaloes with virial masses of $M_\text{vir} \lesssim 10^6\,\Msun$ \citep{Machacek2001,Omukai2001,Agarwal2012}. More massive atomic cooling haloes, on the other hand, become self-shielding so that molecules can form again, eventually leading to star formation \citep{OhHaiman2002}. This means that self-consistent calculations must consider both the radiative transfer and chemical rate equations \citep{Draine1996,Kitayama2004,Wolcott-Green2011,Schauer2015,Schauer2017a}. Furthermore, the shape of the spectral energy distribution affects the relevant photodissociation and photodetachment rates so that it is technically insufficient to parametrize outcomes based only on the strength of the LW radiation field required to suppress $\text{H}_2$ cooling, typically $J_\text{crit} \sim 10^2$--$10^4$ in conventional units of $10^{-21}\,\text{erg/s/cm}^2\text{Hz/sr}$ \citep{Omukai2001,Agarwal2015}. Nonetheless, the first simulations with cosmological initial conditions presumed prior destruction of $\text{H}_2$ via a super-critical LW flux, demonstrating that the inflow would continue on its near-isothermal track \citep{BrommLoeb2003}. The resulting free-fall collapse of the atomically cooling gas could then produce massive black holes directly. Subsequently, the DCBH model has received increased attention with a particular focus on the conditions of collapse, overcoming the angular momentum barrier, consequences of disc (in)stability, supersonic turbulence properties, formation of a radiation pressure-supported supermassive star, accretion rates, feedback processes, and final DCBH mass function and abundance estimates in the cosmological context \citep{Begelman2006,Lodato2006,Lodato2007,Regan2009,Choi2013}.

Semi-analytic models and hydrodynamics simulations suggest that strong LW radiation fields can be supplied directly from neighboring galaxies in close proximity to the DCBH formation site \citep{DijkstraHaiman2008,Agarwal2014,Visbal2014,Regan2017}. This scenario is most plausible for a pair of coincidentally synchronized galaxies in a ``Goldilocks zone'' of separation and timing, to maintain sufficient flux while avoiding ram-pressure stripping or external metal pollution. However, statistical searches for neighboring halo DCBH candidates in large-volume cosmological simulations find much fewer than expected \citep{Habouzit2016,Agarwal2018}, and therefore only represents a fraction of the present-day SMBH population. Still, there are several other mechanisms to enable more robust DCBH formation under less extreme LW radiation conditions. For example, it has been suggested that trapped Ly$\alpha$ cooling radiation during the initial collapse may enhance the assembly of DCBHs by facilitating the photodetachment of $\text{H}^{-}$ ions, which are chemical precursors to $\text{H}_2$ \citep{Agarwal2015,JohnsonDijkstra2017}, or by stiffening the thermodynamic equation of state for the gas directly \citep{Spaans2006,GeWise2017}. Alternatively, strongly amplified magnetic fields and turbulence can provide additional pressure support on small scales to increase the viability of the DCBH scenario \citep{Grete2019}.

\begin{figure}
\centering
\includegraphics[width=0.65\textwidth]{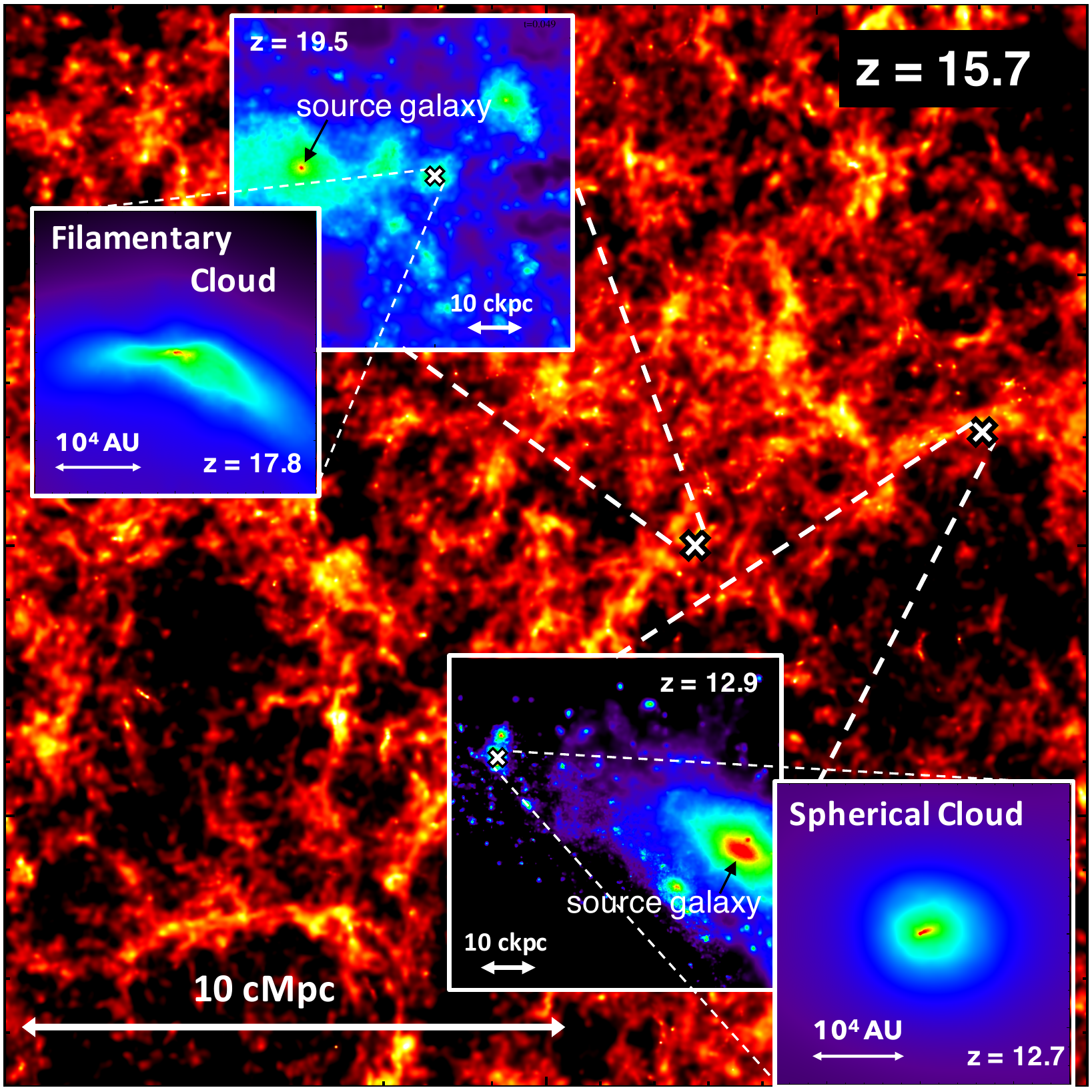}
\caption{Images of filamentary and spherical clouds taken from a large-scale cosmological simulation with a box size of $20\,h^{-1}\,\text{Mpc}$ on each side. The zoom-in regions are potential sites for massive seed black hole formation with central densities above $10^8\,\text{cm}^{-3}$ during the early run-away collapse phase. Follow-up radiation-hydrodynamics simulations capture the subsequent long-term evolution of the protostellar accretion phase for $\sim0.1\,\text{Myr}$ \citep[adopted from][]{Chon2018}.} \label{fig4:Chon2018}
\end{figure}

More generally, shock heating due to supersonic accretion along cosmological filaments, high baryonic streaming velocities relative to dark matter, or violent merger histories may suppress star formation in route to DCBH formation \citep{Schauer2017b,Inayoshi2018}. In particular, streaming velocities have been shown to play an important role in suppressing star formation in minihaloes \citep{Greif2012,Stacy2012,Fialkov2012,Schauer2019}, and may be a powerful mechanism to enhance the formation of massive seed black holes \citep{Hirano2017}. Numerical models may include any of these effects from the initial collapse through the accretion and feedback phases with high accuracy \citep{Becerra2018}. However, fully coupled, radiation hydrodynamics simulations hold the key for further progress and already indicate profound differences from an adiabatic equation of state during both early and late stages affecting the growth of the central structure and anisotropic escape of ionizing photons in 3D geometry \citep{Luo2018,Ardaneh2018}. By utilizing such simulations it was recently argued that the dynamics of structure formation, rather than a critical LW flux, is the main driver of the formation of DCBHs \citep{Wise2019}. This paradigm, or similar variations \citep{Mayer2019}, is attractive because massive black hole seeds can form with less stringent metallicity requirements, implying that they may be much more common than previously considered in overdense regions of the early Universe.

\subsection{Runaway mergers in dense star clusters}
\label{sec:clusters}
The final SMBH formation scenario to be discussed is that of dense clusters undergoing runaway collapse. Such an origin represents an intermediate case between monolithic collapse of DCBHs and the staggering growth required for isolated stellar remnant black holes. In fact, it can be insightful to consider dense cluster seeding as ``failed'' DCBHs with fragmentation initially suppressed to facilitate a high Jeans mass that ultimately gives way to instabilities during later stages. Optimistically, with a ubiquitous supply of cold gas effectively trapping the accretion radiation, a $\sim10\,\Msun$ black hole seed subjected to random motions through such a cluster may initiate supra-exponential growth over a dynamical timescale \citep{AlexanderNatarajan2014}. However, this scenario is extremely difficult to model self-consistently in the context of \textit{ab initio} cosmological simulations. This goal is important, because large-scale effects such as tidal forces within hierarchical structure formation can induce higher stellar multiplicities \citep[][see Fig.~\ref{fig4:Chon2018}]{Chon2018}. It may be the case that massive primordial star clusters experience high gas accretion rates, thereby enhancing the effective stellar radii and collisional cross-sections, thus achieving the conditions required for runaway mergers \citep{Boekholt2018}. Considering the final runaway dynamics may also involve higher order N-body interactions \citep{Katz2015}. In any case, the diversity of conditions in the early Universe may allow for the concurrent formation of supermassive stars and globular clusters \citep{Gieles2018}.

It is also intriguing to establish connections between the present-day population of nuclear SMBHs and the less understood census of $\lesssim 10^6\,\Msun$ intermediate mass black holes. Even if dense clusters fail to explain the first SMBHs, they represent a generic formation mechanism for other abundant objects in the Universe, including the progenitors of globular clusters \citep{LiGnedin2017,Boylan2018,LiGnedin2019}. Encouragingly, observations reveal massive $\gtrsim 10^5\,\Msun$ super star clusters with sizes of only a few parsecs in local starbursts and lensed or dusty star-forming galaxies at high redshifts \citep{JohnsonKobulnicky2003,Clark2005,Vanzella2017}. Super star clusters are characterized by extreme star formation rate densities, and it is believed that mechanical feedback and radiation pressure play nontrivial roles in the disruption of the natal giant molecular clouds \citep{Murray2010}. Three-dimensional radiation hydrodynamics simulations of super star clusters are indicative of the conditions for massive black hole nurseries, but the simulation complexity, including resolution requirements, isolated environment, and additional physics, renders the SMBH connection speculative \citep{Skinner2015,Tsang2018}. Furthermore, if super star clusters represent an evolutionary pathway to present-day globular clusters, by the time a massive black hole is detectable, it may be difficult to discern a cluster origin given the increasingly aging and mixed stellar populations.

\section{Impact on galaxies and the early IGM}
\label{sec:impact}
Although nuclear black holes typically comprise only a small fraction of the available mass in galaxies, their efficient conversion of rest mass to radiation provides a powerful source of feedback that drives winds, regulates star formation, and contributes to the reionization of the Universe. At $z \lesssim 10$ the differences arising from the distinct SMBH formation scenarios are progressively erased as the host galaxies experience minor and major mergers \citep{Valiante2018}. Therefore, in this section we briefly discuss the importance of SMBHs in the high-redshift Universe without a particular emphasis on their precise origins. Overall, feedback from active galactic nuclei (AGN) is an essential ingredient in galaxy formation models to couple small and large scales, for example, accretion discs, relativistic jets, co-evolution with stars, quenching massive galaxies, and heating the IGM.

\begin{figure}
\centering
\includegraphics[width=0.64\textwidth]{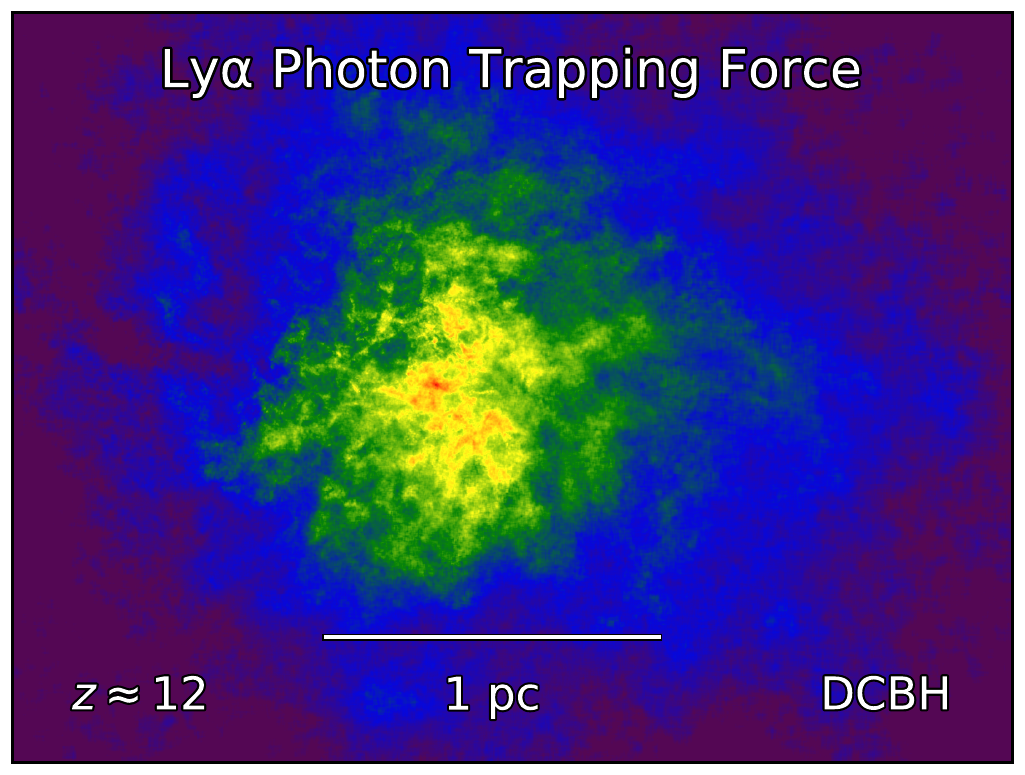}
\caption{Illustration of Ly$\alpha$ photon trapping in a DCBH assembly environment. The central regions (red--green) experience strong turbulent accelerations of order $a_{\text{Ly}\alpha} \gtrsim 10\,\text{km/s/kyr}$ due to Ly$\alpha$ radiative feedback. Although this problem requires fully-coupled radiation hydrodynamics, post-processing radiative transfer simulations already indicate that Ly$\alpha$ radiation pressure can be dynamically important to first aid DCBH formation and then provide self-regulating feedback to limit the final mass \citep[adapted from][]{SmithBecerra2017}.} \label{fig5:SmithBecerra2017}
\end{figure}

\subsection{Radiative feedback}
\label{sec:feedback}
At the extreme end of SMBH-accretion powered AGN activity are the quasars, with distinguishing characteristics, as follows: (\textit{i}) the luminosity can be orders of magnitude brighter than their host galaxy, (\textit{ii}) the central engines are extremely small and unresolved in observations, (\textit{iii}) the significant variability of the observed flux over short timescales, and (\textit{iv}) the non-thermal spectral energy distributions, which often follow a power law out to high energies with strong X-ray emission. The non-thermal continuum is generated by bremsstrahlung radiation due to the acceleration of electrons by ions, Compton scattering of photons by free electrons with the corresponding frequency shifting, and synchrotron radiation due to electrons accelerating in a magnetic field. Additionally, the accretion disc radiates broadband emission with a peak in the UV, corresponding to a characteristic blackbody temperature for the Eddington luminosity near the event horizon of $T_\text{Edd} \approx 5 \times 10^5\,\text{K}\,(M_\bullet/10^8\,\Msun)^{-1/4}$ \citep{Rees1984}. The canonical AGN morphology is that of an obscuring molecular torus surrounding a compact accretion disc, with fast moving ($\gtrsim 1000\,\text{km/s}$) broad-line-region clouds out to tens of parsecs transitioning to narrow-line-region clouds at larger distances. The relative importance of these features including the presence of a relativistic jet along with the observed orientation provides a unification scheme for AGN classifications \citep{Urry1995}.

Cosmological hydrodynamics simulations often do not resolve the relevant scales for SMBH growth. The approximate models are based on inefficient thermal feedback at high accretion rates as the dominant growth channel, until a self-regulating kinetic feedback mode maintains lower accretion rates with SMBH mergers becoming the main channel for subsequent mass growth \citep{Steinborn2018,Weinberger2018}. In the first galaxies, radiation pressure from the black hole along with concurrent star formation and supernovae are likely to have had a substantial impact on their surrounding environments \citep{Jeon2012}. Lower mass minihaloes with shallower gravitational potential wells ($M_\text{vir} \sim 10^5$--$10^7\,\Msun$) would have been especially susceptible to radiative feedback, which potentially depletes the reservoir of gas needed to fuel black hole growth \citep{Wise2012}. A particularly interesting source of feedback arises from the resonant scattering of Ly$\alpha$ photons (see Fig.~\ref{fig5:SmithBecerra2017}). In 1D simulations, a radial outflow forms in response to the central source ionizing and heating the gas, with Ly$\alpha$ photons significantly enhancing such radiation-driven winds \citep{DijkstraLoeb2008,SmithRHD2017}. An approximate model for Ly$\alpha$ pressure was recently developed for 3D hydrodynamics simulations \citep{Kimm2018}, which demonstrated an impact on metal-poor dwarf galaxies by regulating the dynamics of star-forming clouds before the onset of supernova explosions. Future cosmological simulations with an accurate and efficient treatment of Ly$\alpha$ feedback will provide new perspectives on the growth of the first stars and SMBHs, cold gas accretion flows, and turbulence in high-redshift galaxies \citep{SmithFIRE2019,SmithRDDMC2018}.

\begin{figure}
\centering
\includegraphics[width=0.65\textwidth]{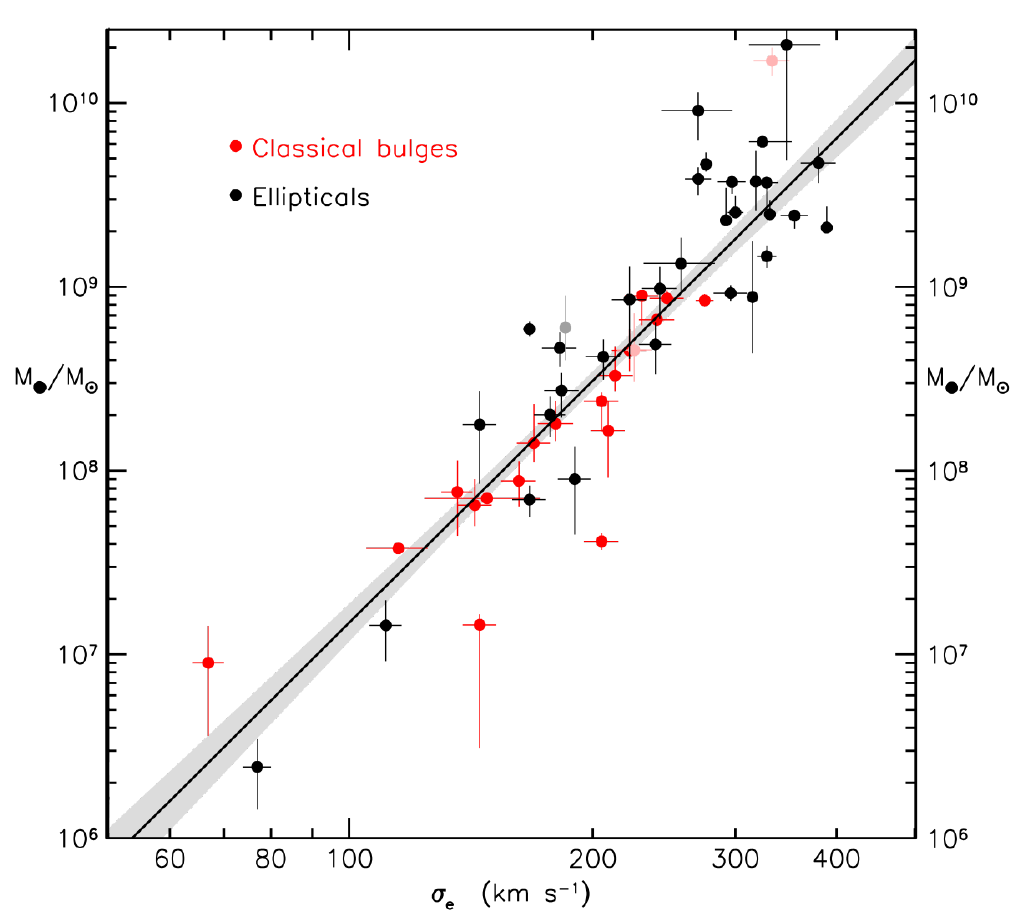}
\caption{The $M_\bullet$--$\sigma_\star$ relation is based on the remarkable correlation between the observed central black hole mass $M_\bullet$ and stellar velocity dispersion $\sigma_\star$ in the bulge of galaxies, suggesting the existence of a feedback mechanism to couple the growth of nuclear SMBHs to the mass of the galaxy. The relation extends across many orders of magnitude in mass and holds for both classical bulges and elliptical galaxies (morphological classifications). The line is a symmetric least-squares fit to the data with the 1$\sigma$ range shown as shaded regions. \citep[adapted from][]{Kormendy2013}.} \label{fig6:Kormendy2013}
\end{figure}

\subsection{Co-evolution with stars}
\label{sec:coevolution}
The existence of SMBHs at the centre of almost every galaxy provides strong evidence for a generic growth mechanism alongside that of the host galaxy. Indeed, a remarkable correlation has been established between the central black hole mass $M_\bullet$ and stellar velocity dispersion $\sigma_\star$ in the bulge (central nucleus) of galaxies: $M_\bullet = 0.310^{+0.037}_{-0.033} \times 10^9\,\Msun\,[\sigma_\star/(200\,\text{km/s})]^{4.38\pm0.29}$ \citep[see Fig.~\ref{fig6:Kormendy2013},][]{Silk1998,Ferrarese2000,Kormendy2013}. The so-called $M_\bullet$--$\sigma_\star$ relation strongly suggests the existence of a feedback mechanism to couple the growth of nuclear SMBHs to the mass of the galaxy. In fact, the co-evolution could originate as a consequence of the SMBH formation mechanism. For example, the collapse of primordial gas clouds with successful massive seed black holes would be followed by star formation in the spheroidal bulge \citep{Barrow2018}. The virialization balance between the gravitational potential and kinetic energies produce the observed stellar velocity dispersion. As black hole accretion continues, the intense radiative feedback periodically regulates the gas supply to the bulge and the feeding of the black hole. On the other hand, SMBHs do not correlate with galaxy-scale discs and are only weakly related to disc-grown ``pseudobulges'' or properties of the host dark matter haloes. This suggests multiple regimes of black hole feedback including effects from the local environment and galaxy mergers that complicate the high-level picture described above.

Another aspect of co-evolution is the existence of nuclear star clusters and nuclear discs that co-exist with SMBHs at the centers of low- and intermediate-mass spheroids \citep{Wehner2006}. Interestingly, massive elliptical galaxies do not contain nuclear star clusters, presumably as a result of dynamical friction and evaporation during hierarchical merging. It is also possible that the most extreme SMBH formation sites were violent enough to also preclude efficient nuclear star formation. However, at decreasing spheroid mass, the nuclear star cluster eventually dominates over the SMBH component, consistent with an absence of a central black hole in globular clusters \citep{Graham2009}. When both a SMBH and nuclear star cluster exist, it may make sense to combine the masses for an $(M_\bullet + M_\text{NSC})$--$\sigma_\star$ relation with less intrinsic scatter \citep{Nayakshin2009,Graham2011}. Inferences about high-redshift low-luminosity AGN can be affected by the presence of nuclear star clusters due to competing feedback, biased velocity dispersions, and errors in mass estimates, while at the same time potentially providing clues about the evolution of the first SMBHs. Recent observations indicate that the majority of $z \sim 6$ quasars are more massive than expected from the local $M_\bullet$--$M_\text{halo}$ relation, with a third of the known population by factors $\gtrsim 10^2$ \citep{Shimasaku2019}. These numbers are likely affected by selection bias and unknown completeness, but clearly indicate that the growth of the most massive high-redshift SMBHs greatly precedes that of hosting haloes, implying that the $M_\bullet$--$\sigma_\star$ relation is not initially established via symbiotic growth at early times \citep{Volonteri2012}.

\subsection{Mini-quasars}
\label{sec:quasars}
Radiation from the first stars and galaxies initiated the dramatic phase transition known as the ``epoch of reionization'' marking an end to the cosmic dark ages within the first billion years after the Big Bang. While it is generally accepted that stars contributed a significant fraction of the ionizing photon budget, the role of X-ray sources including low-luminosity mini-quasars to the reionization process is tentatively constrained to be subdominant \citep{HaardtMadau2012,MadauHaardt2015}. The X-rays from early AGN heat the IGM above the CMB temperature, rendering the 21\,cm hyperfine-structure line of hydrogen visible in emission at $z \lesssim 15$ \citep{Furlanetto2006,PritchardLoeb2012}. Both the neutral hydrogen fraction and 21\,cm spin temperature depend on the spectral shape of the ionizing radiation, giving rise to either a more patchy or homogeneous IGM, with soft X-ray sources being more efficient at heating and ionizing than hard X-ray sources \citep{Fialkov2015}. There seems to be mounting evidence for so-called ``late'' reionization in which most of the volume of the Universe is rapidly ionized around $z \sim 7$--$8$ as opposed to the previous idea that reionization was well under way by $z \sim 10$. This understanding comes from: (\textit{i}) the \textit{Planck} measurement of a low optical depth for electron scattering of CMB photons \citep[$\tau_e = 0.054 \pm 0.007$,][]{Planck2018}, (\textit{ii}) the declining fraction of Ly$\alpha$ emitters among the galaxy population at $z \gtrsim 6$ \citep{Stark2011,Schenker2014,Mason2019}, (\textit{iii}) the imprint of neutral hydrogen in the IGM as a damping wing absorption feature on the spectrum of high-redshift quasars \citep{Simcoe2012,Davies2018}, and (\textit{iv}) the thermal history of the IGM \citep{Onorbe2017,Kulkarni2018}. Models with dominant AGN contributions are typically disfavored due to constraints of the quasar luminosity function at $z \lesssim 5$ and delayed helium reionization at $z \lesssim 3$ \citep{McGreer2018,Puchwein2019}. On the other hand, Ly$\alpha$ forest observations also provide a measurement of the integrated time that galaxies shine as AGN. Such a procedure has identified a population of high-redshift quasars with very small proximity zones, indicating active lifetimes of less than $10^4$\,years in support of massive seeding and rapid growth scenarios \citep{Eilers2017,Eilers2018}. We anticipate a more complete understanding of the epoch of reionization from complementary upcoming facilities, including the SKA and \textit{JWST}, which will provide essential clues about the formation and evolution of SMBHs in the high-redshift Universe.

\begin{figure}
\centering
\includegraphics[width=0.7\textwidth]{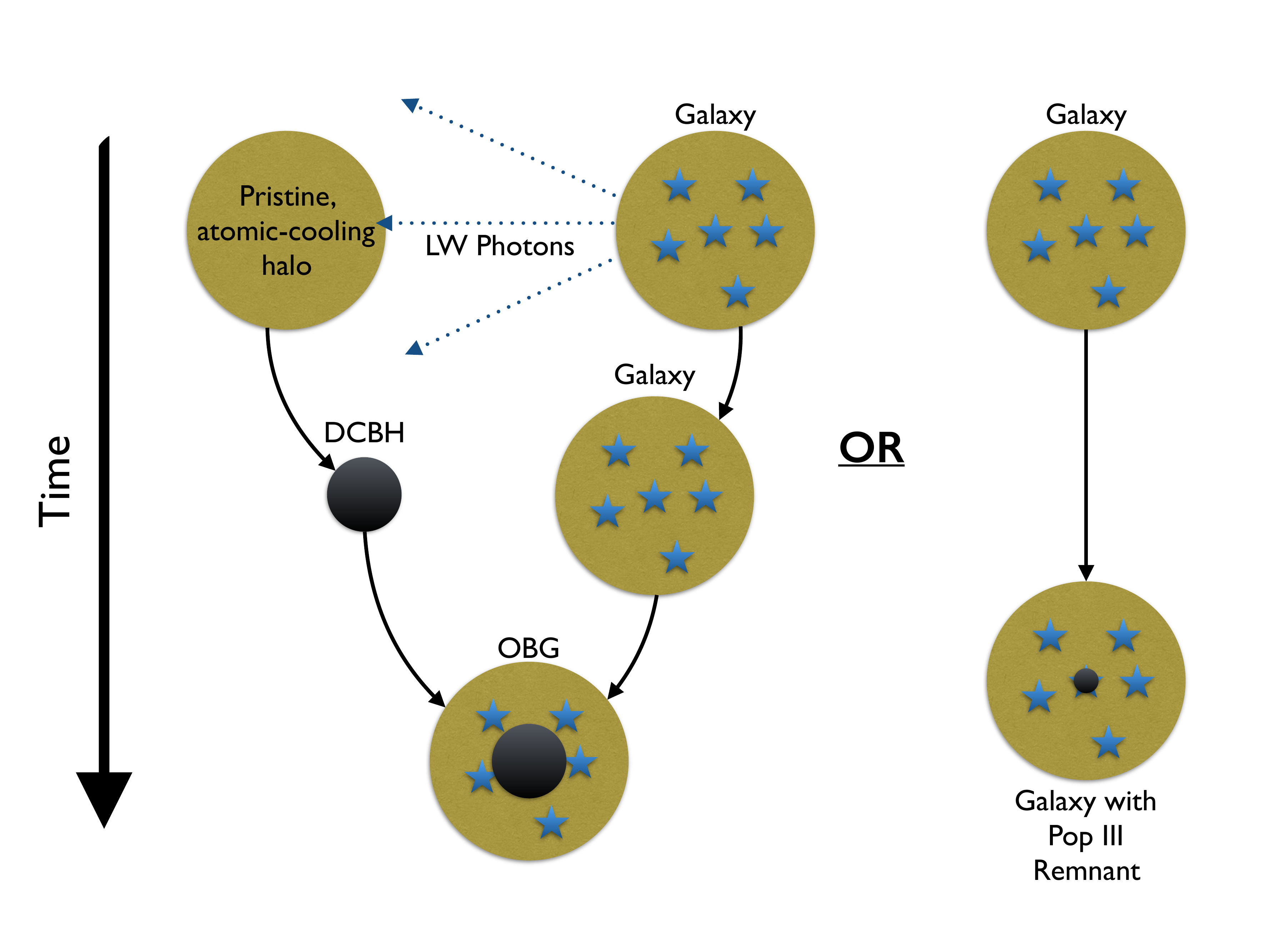}
\caption{Schematic illustration of two seed scenarios in which the black hole is either obese or subdominant. \textit{Left panel:} A DCBH seed ($\sim 10^5\,\Msun$) merges with its parent star forming halo, which provided the LW radiation that facilitated its assembly. \textit{Right panel:} A Pop~III stellar remnant black hole ($\sim 100\,\Msun$) grows within a previously star forming galaxy. Multi-wavelength observations and predictions will play an important role in distinguishing formation scenarios in the era of next-generation telescopes, such as the \textit{JWST} \citep[adopted from][]{Natarajan2017}.} \label{fig7:Natarajan2017}
\end{figure}

\section{Detection}
\label{sec:detection}
Our emerging understanding of structure formation, within the overall framework of standard cosmology, clearly indicates that black holes play an important role already early on in cosmic history. It is therefore important to develop a network of empirical probes to test the theoretical picture. There are a number of promising avenues to accomplish this in the upcoming decade and beyond with a suite of next-generation observational facilities. The upcoming surveys will cover a broad range of wavelengths, and crucially will have a multi-messenger character. In briefly summarizing some of those approaches, it is useful to distinguish direct (``in-situ'') observations at high redshifts from ``fossil'' probes, where signatures of the first black holes are captured at later cosmic times.

\subsection{In-situ methods}
\label{sec:in-situ}
\begin{figure}
\centering
\includegraphics[width=0.7\textwidth]{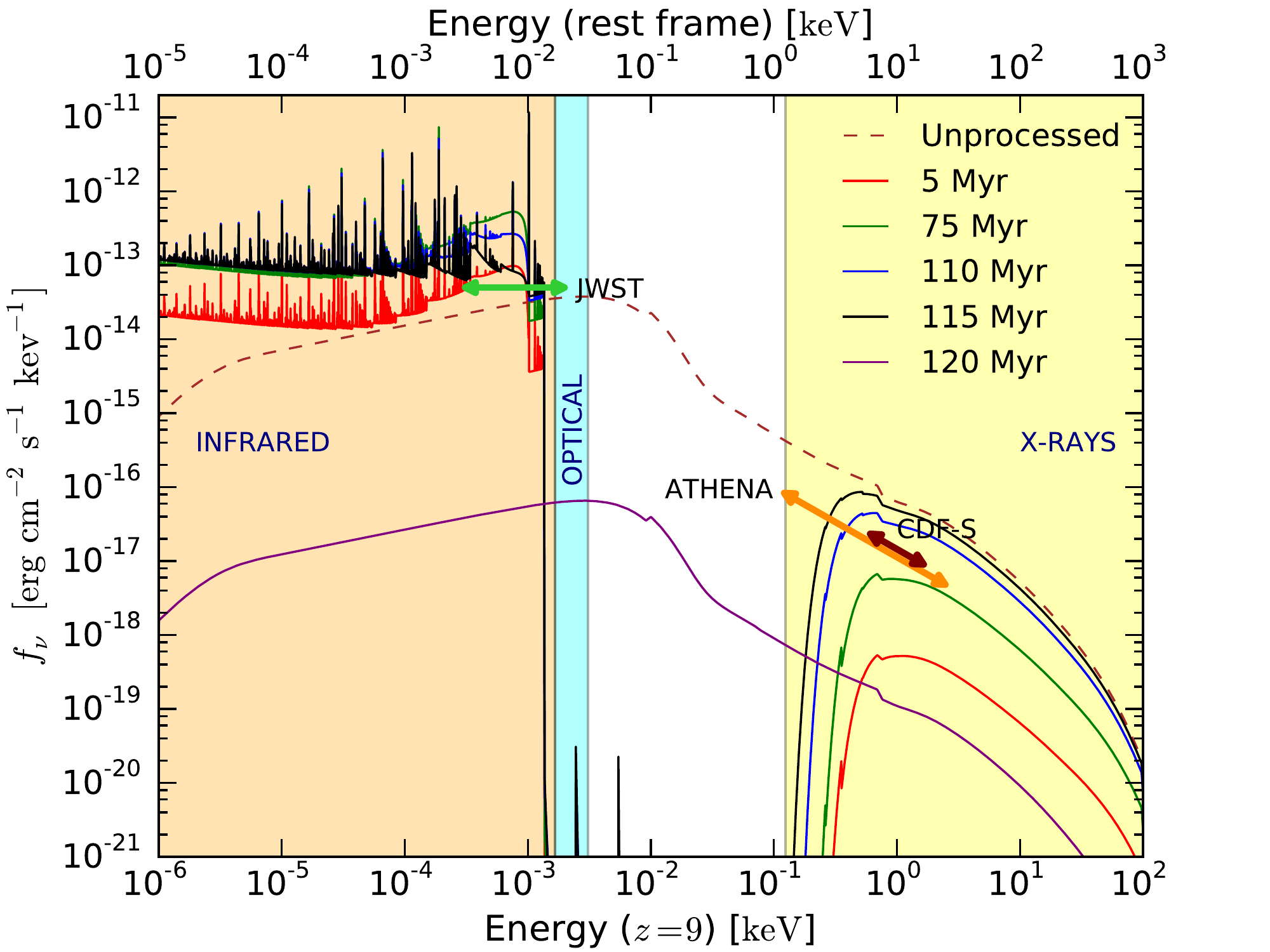}
\caption{Emerging spectrum from a DCBH host system at $z\sim 10$. It is evident how the unprocessed spectrum (\textit{dashed line}) is reprocessed by the high column-density in the envelope into the characteristic `double-humped' shape during early DCBH evolution. Eventually, after $\sim 120\,\text{Myr}$, the previous accretion has depleted the high-density gas, so that the central spectral gap is filled in again. The key wavebands for capturing this spectral signature are the near-infrared with the \textit{JWST}, and the X-ray band with \textit{ATHENA} (sensitivity thresholds are indicated) \citep[adopted from][]{PacucciDCBH2015}.}
\label{fig8:Pacucci2015}
\end{figure}

A key target for high-redshift observations is to look for evidence of DCBH seeds. Since these objects are ``born massive'', they would stand out in a number of ways. First, they would initially present a pure black hole (AGN) signature, with a weak, or non-existent, stellar component in the overall source spectral energy distribution (SED). Such an early AGN-dominated phase has been termed ``obese'', and has been suggested as a telltale DCBH feature \citep[][see Fig.~\ref{fig7:Natarajan2017}]{Natarajan2017}. Second, the early growth phase of a DCBH is predicted to be heavily obscured, where a massive gaseous envelope is rapidly accreted onto the central source \citep{Yue2013,PacucciDCBH2015}. The high column density in neutral hydrogen will efficiently absorb UV photons, produced in the inner accretion flow, above the hydrogen-ionization threshold. Those photons will be reprocessed into lower-energy infrared radiation, whereas any non-thermal X-ray photons can largely escape unobscured. The resulting radiative transport thus imprints a characteristic ``double-humped'' spectral profile, with a peak in the near- and mid-infrared, as well as $\sim\,\text{keV}$ X-ray bands (see Fig.~\ref{fig8:Pacucci2015}). Deeply embedded inside the massive accretion flow, an initially ultra-compact ionized region is developing \citep{SmithBecerra2017}. At some stage, this growing ionized region will break out of the embedding cocoon, thus effectively terminating the rapid DCBH assembly process. As a consequence of the dynamics, photons will experience an increased opacity to scattering from the abundant free electrons. The flow, in effect, becomes ``Compton thick'', or more precisely ``Thomson thick'', such that $\tau_e \gtrsim 1$ (see footnote~\ref{footnote:optical_depth}). After a few $10\,\text{Myr}$, this initial, highly obscured and possibly Compton-thick, evolutionary phase will end, and the emerging radiation will transition into a standard AGN signature. The precise duration of this initial stage is uncertain, thus impacting any assessment for the likelihood of catching a DCBH in its early characteristic phase.

To hunt for the DCBH signature, \textit{JWST} observations in the near- and mid-infrared will have the required sensitivity for broad-band photometry with its NIRCam and MIRI instruments. Given the uncertainty in predicting source number densities, such deep \textit{JWST} searches are ideally complemented by wide-field surveys with the \textit{Wide Field Infrared Survey Telescope} (\textit{WFIRST}), further ahead in time. Similarly, it will be crucial to mirror the infrared observations with simultaneous X-ray campaigns. The latter are ideal targets for the planned \textit{Advanced Telescope for High ENergy Astrophysics} (\textit{ATHENA}) or \textit{Lynx} X-ray missions. In terms of such correlated signatures in the infrared and X-ray bands, there is an independent strategy to access it, by focusing on the aggregate signal imprinted in the cosmic infrared and X-ray backgrounds \citep{Kashlinsky2018}. Indeed, there are tantalizing hints for a correlated signal in the two backgrounds, with statistical properties that can be attributed to an unresolved population of high-redshift sources that emit \textit{both} in soft-UV photons and hard X-rays. After cosmological redshifting, those emissions would be deposited in the present-day cosmic backgrounds \citep{Cappelluti2013}. It is possible that DCBH sources are, at least partially, responsible for the inferred signature in these cosmic backgrounds \citep{Helgason2016}.

\begin{figure}
\centering
\includegraphics[width=0.7\textwidth]{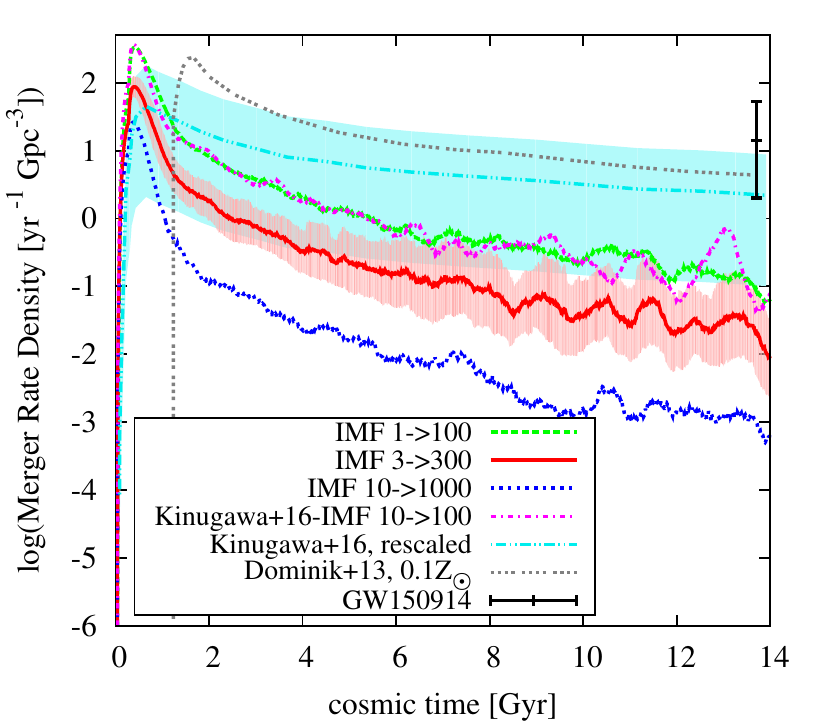}
\caption{Compact binary merger rate for Pop~III sources, plotted as a function of cosmic time, for different assumptions on the initial mass function. The thick red line with a shaded error region represents the preferred model. Similar curves from other studies are reproduced for reference \citep{Dominik2013,Kinugawa2016}. The black data point with error bars (\textit{upper right}) is the estimated binary black hole merger rate as derived from the initial LIGO detection. As can be seen, Pop~III sources may contribute of order 1\% of the full LIGO sample \citep[adopted from][]{Hartwig2016}.} \label{fig9:Hartwig2016}
\end{figure}

\subsection{Fossil methods}
\label{sec:fossils}
The gravitational-wave signature from merging black hole binaries offers an exciting new window into the cosmic history of black hole formation and growth. With advanced LIGO, the black hole remnants from massive Pop~III progenitor stars may be detectable as a subset of the rapidly growing overall sample \citep{Hartwig2016,Inayoshi2017,Pacucci2017}. Any predictions for the expected Pop~III binary black hole merger rates as a function of redshift are subject to uncertainties in the input physics, such as the Pop~III initial mass function and binary stellar evolution at zero metallicity \citep{Belczynski2014}. Approximate calculations, however, indicate that of order 1\% of the total LIGO binary black hole sample may originate in Pop~III stellar systems, with about one expected merger per year at LIGO's final design sensitivity (see Fig.~\ref{fig9:Hartwig2016}). A key challenge is to disentangle any Pop~III gravitational-wave signal from the dominant population of metal-enriched progenitors, given that the currently observed black hole mass range of a few tens of solar masses in itself is not uniquely identifying Pop~III. Such a unique Pop~III signature would be provided by black hole masses of $\sim 300\,\Msun$, given that metal-enriched massive stars are not likely to reach this extreme mass range. In the latter case, vigorous stellar winds would trigger strong mass loss, which may largely be absent for Pop~III stars.

To reach the supermassive regime with gravitational-wave observations, we will have to wait for the Laser Interferometer Space Antenna (LISA) to fly in the 2030s. This space-based observatory with its million-km baselines is sensitive to frequencies of $\sim 0.1$\,mHz--$0.1$\,Hz, thus extending the LIGO band to much longer wavelengths, where the signature of the first SMBHs is expected \citep{VolonteriBellovary2012}. Specifically, the LISA sensitivity should enable the detection of black hole mergers with masses $\sim 10^3$--$10^6\,\Msun$, all the way to redshifts of $z\sim 20$. One of the key questions now is how to distinguish between the two main scenarios to seed black hole growth, light stellar seeds and heavy DCBH ones \citep{RicarteNatarajan2018}. Through a combination of X-ray observations at high-redshift (see above) and the statistical analysis of the gravitational-wave events, as detected by LIGO and LISA, we can build up the mass spectrum of the first black holes. In principle, this determination should allow us to identify the dominant black hole formation channel in the pre-reionization Universe.

\section{Outlook}
\label{sec:outlook}
Our emerging picture of early cosmological structure formation contains the intriguing message that black holes played an important role early on. Indeed, the first black holes provide us with a multitude of probes into the formative initial period of cosmic history. Furthermore, it has become evident that they were key players, actively shaping early cosmic evolution through their feedback effects on the primordial gas. This feedback in turn comprises a large range of scales, from the immediate surroundings of the compact object, to the interstellar medium in the host haloes, reaching all the way into the large-scale cosmic web. The Universe throughout cosmic times contains black holes over a wide range of masses, but it is remarkable that black hole formation in the early Universe may have received a `headstart', biasing any formation channels to higher masses. Such possible bias would mirror the predicted top-heavy nature of the first (Pop~III) stars.

The first black holes also add powerful new avenues for observational astronomy. At the supermassive scale, the \textit{JWST} may detect the initial, heavily-obscured, formation phase of DCBHs, as well as their subsequent accretion and growth phase. Crucially, these sources provide us with portals into multi-wavelength and multi-messenger surveys. As to the former, the spectral energy distribution of the black hole emission is extended into the X-ray regime, opening up the high-energy window both for individual sources and their combined contribution to the cosmic X-ray background. At the extreme energy end, a subset of the first black holes is expected to be born in rare gamma-ray burst events \citep{Toma2016}, accessible to future all-sky gamma-ray burst missions, such as the Transient High-Energy Sky and Early Universe Surveyor (THESEUS). Additionally, 21\,cm cosmology measurements with the SKA will provide key insights via constraints on the cold primordial gas in the cosmic dark ages and the epoch of reionization. Even more exciting, as part of the developing field of multi-messenger astronomy, are the prospects to link the electromagnetic emission from the first black holes with their gravitational-wave signature, and possibly even with the search for ultra-energetic neutrinos. Given the confluence of breakthroughs in the theoretical modelling as well as in observational capabilities, the upcoming decade promises to be a golden age of discovery for the first supermassive objects in the Universe.

\newpage

\section*{Acknowledgements}
We thank Kohei Inayoshi, Anna Schauer, and the anonymous referees for very helpful comments on the draft. Support for Program number HST-HF2-51421.001-A was provided by NASA through a grant from the Space Telescope Science Institute, which is operated by the Association of Universities for Research in Astronomy, Incorporated, under NASA contract NAS5-26555. VB acknowledges support from NSF grant AST-1413501.

\section*{Notes on contributors}

\begin{wrapfigure}{l}{0.13\textwidth}
  \vspace{-0.42cm}
  \includegraphics[trim=0in -14in 0in 1in,clip,width=0.14\textwidth]{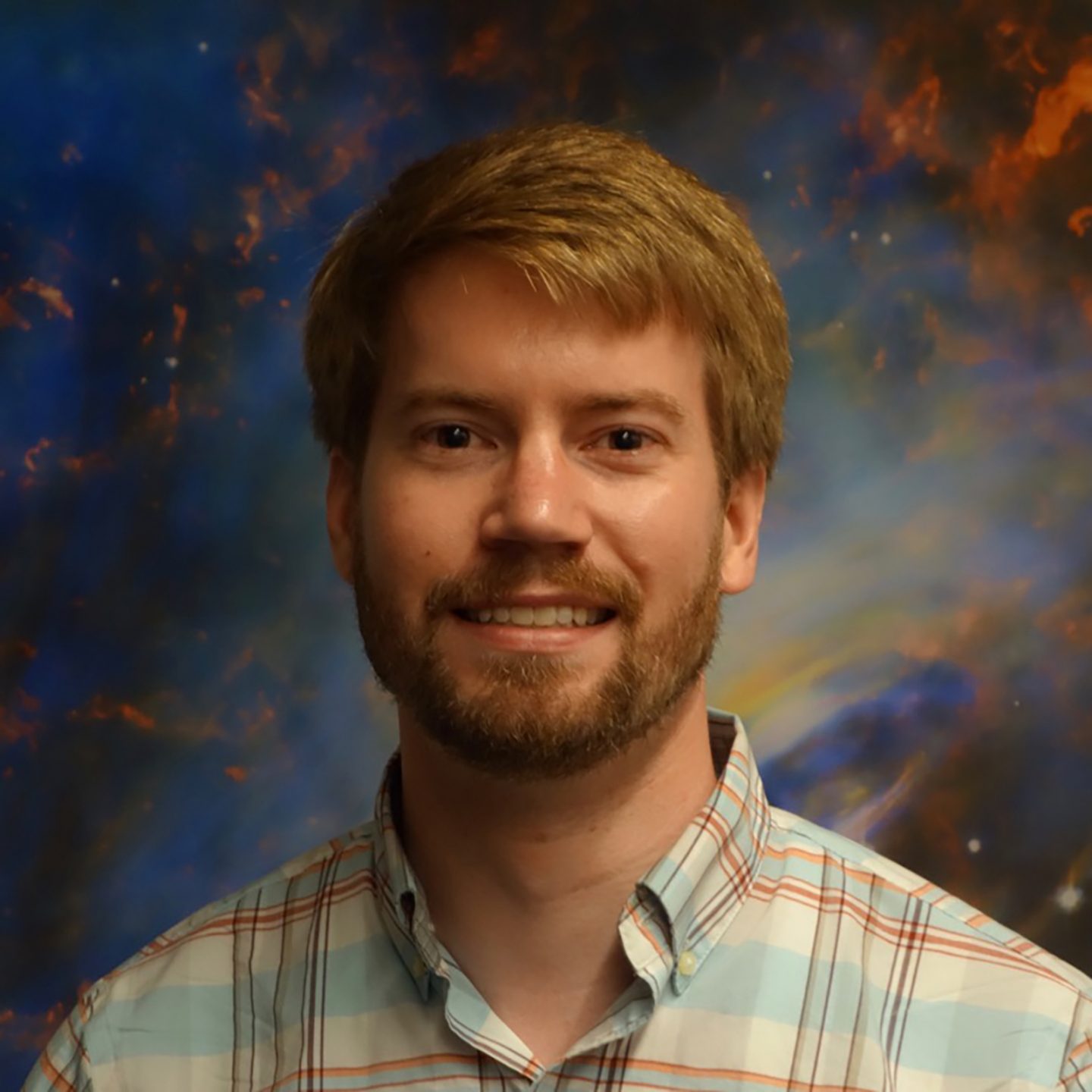}
\end{wrapfigure}
Aaron Smith is currently a NASA Einstein Fellow at the Kavli Institute for Astrophysics and Space Research at the Massachusetts Institute of Technology. Smith recently received his PhD from the University of Texas at Austin where he held a National Science Foundation Graduate Research Fellowship. He primarily works in galaxy formation theory and numerical radiation hydrodynamics.

\vspace{0.5cm}

\begin{wrapfigure}[1]{l}{0.13\textwidth}
  \vspace{-4.05cm}
  \includegraphics[trim=2in 0in 1.5in 0in,clip,width=0.14\textwidth]{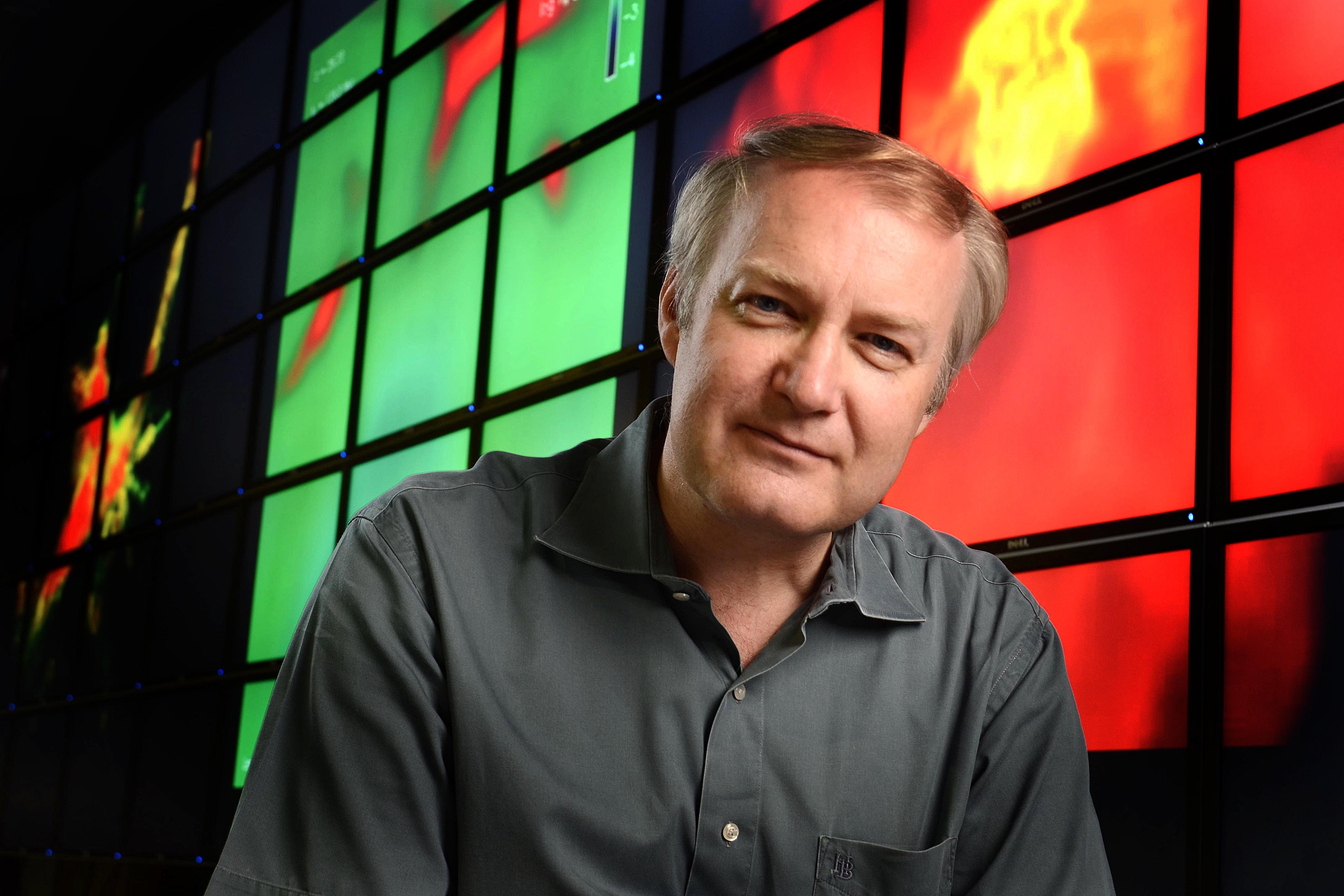}
\end{wrapfigure}
\noindent Volker Bromm is a professor of Astronomy at The University of Texas at Austin. Bromm was educated at Yale University before undertaking postdoctoral research at Cambridge University and the Harvard-Smithsonian Center for Astrophysics. He has worked on numerous problems in the areas of the first stars and galaxies, dark matter theory, and stellar archaeology, and has published several reviews on these topics.

\bibliographystyle{tfnlm}
\bibliography{biblio}

\end{document}